\documentclass[prl,twocolumn,superscriptaddress,showpacs]{revtex4-1}
\usepackage{graphicx,amsmath,amssymb}
\usepackage{color}
\usepackage[normalem]{ulem}	
\begin{document}

\title{Neutron Scattering Study on Spin Dynamics in Superconducting (Tl,Rb)$_{\rm 2}$Fe$_{\rm 4}$Se$_{\rm 5}$}
\author{Songxue Chi}
\affiliation{
Quantum Condensed Matter Division, Oak Ridge National Laboratory, Oak Ridge,
Tennessee 37831, USA
}
\author{Feng Ye}
\affiliation{
Quantum Condensed Matter Division, Oak Ridge National Laboratory, Oak Ridge,
Tennessee 37831, USA }
\affiliation{Department of Physics and Astronomy, University of Kentucky,
Lexington, Kentucky 40506, USA}
\author{Wei Bao}
\email{wbao@ruc.edu.cn}
\affiliation{Department of Physics, Renmin University of China, Beijing
100872, China }
\author{Minghu~Fang}
\author{H.~D.~Wang}
\author{C.~H.~Dong}
\affiliation{
Department of Physics, Zhejiang University, Hangzhou 310027, China}
\author{A.~T.~Savici}
\affiliation{
Neutron Data Analysis and Visualization Division, Oak Ridge National Laboratory, Oak Ridge,
Tennessee 37831, USA
}
\author{G.~E.~Granroth}
\author{M.~B.~Stone}
\affiliation{
Quantum Condensed Matter Division, Oak Ridge National Laboratory, Oak Ridge,
Tennessee 37831, USA
}
\author{R.~S.~Fishman}
\affiliation{
Materials Science and Technology Division, Oak Ridge National Laboratory, Oak Ridge,
Tennessee 37831, USA }

\begin{abstract}
We observed in superconducting (Tl,Rb)$_2$Fe$_4$Se$_5$ spin-wave branches that span an
energy range from 6.5 to 209 meV. 
Spin dynamics are successfully
described by a Heisenberg localized spin model whose dominant in-plane interactions include only 
the nearest ($J_1$ and $J_1^{\prime}$) and next nearest neighbor ($J_2$ and $J_2^{\prime}$)
exchange terms within and between the tetramer spin blocks, respectively. 
These experimentally determined exchange constants would crucially constrain the theoretical 
viewpoints on magnetism and superconductivity in the Fe-based materials.
\end{abstract}

\pacs{74.25.Ha, 74.70.-b, 78.70.Nx}

\maketitle

One astonishing aspect of the recently discovered $T_c\sim 30$ K iron 
selenide superconductors \cite{C122924,C123637,C125236,C125525,D010462}
is the coexistence of a large magnetic moment (3.3$\mu_B$/Fe) and high transition-temperature 
($T_N\approx 470$-560 K) antiferromagnetic order \cite{D020830,D022882}. 
Different from all other families of the Fe-based superconductors, 
the new iron selenide superconductors consist of Fe plates with 
highly ordered $\sqrt{5}\times\sqrt{5}$ vacancy superstructure \cite{D020830,D014882,D022882}.
Samples with less developed $\sqrt{5}\times\sqrt{5}$ order are 
known to {\em not} be superconducting either when their compositions  
are close to \cite{D020488} or deviate significantly from \cite{D023674} the ideal $A_2$Fe$_4$Se$_5$ formulas. 
In the latter case, an additional phase of orthorhombic vacancy order with a $\sqrt{2}\times\sqrt{2}$
or $2\sqrt{2}\times\sqrt{2}$ unit cell also exists at intermediate temperatures.
The large-moment block antiferromagnetic order developed on the vacancy 
ordered Fe lattice [Fig.~1(a-b)] exists in superconducting as well as 
insulating samples \cite{D023674}.

The perfect $\sqrt{5}\times\sqrt{5}$ vacancy order demands one vacancy 
per five Fe ions and the charge neutrality enforces a proportional number
of intercalating $A$ ions. While superconductivity can tolerate, or even
requires, a small composition deviation from the ideal $A_2$Fe$_4$Se$_5$ to dope charges, the  
crystal structure has to deform, as expected in any non-stoichiometric samples, 
such that the ``vacant'' Fe1 site is found to be occupied by a few percent of Fe 
on average \cite{D020830,D014882} and fine-scale structural 
variation \cite{D012059,D030059} and phase-separation \cite{D070412} 
are observed in superconducting samples. 
Nonetheless, the block antiferromagnetic order
not only coexists with superconductivity in the same sample
\cite{D020830,D022882,D011873,D030059}, 
but also the interaction between these two long-range ordered states reveals itself 
in an anomalous magnetic order parameter near $T_c$ \cite{D020830,D022882,D062706}.
Irrespective of the current theoretical debate on the role played by spin excitations in 
forming the superconducting Cooper pairs \cite{D063712,D031902,E025563}, it is important to investigate spin dynamics experimentally
in the new iron selenide superconductors to bound the discussion.

\begin{figure}[b!]
\label{fig1}
\includegraphics [width=.9\columnwidth]{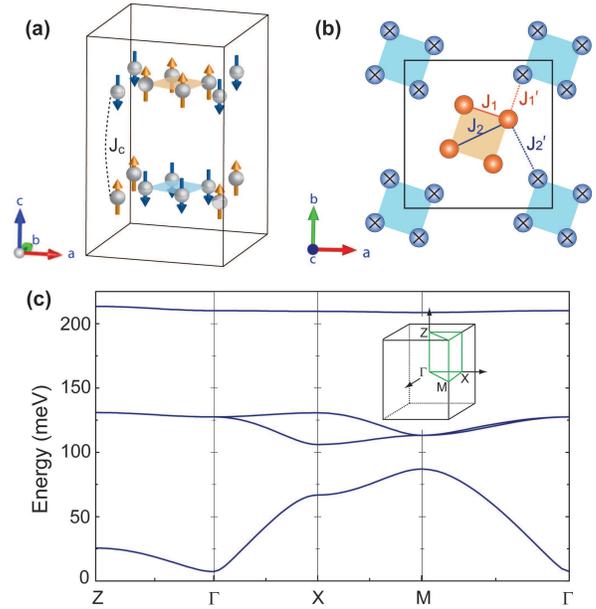}
\caption{(color online) (a) Schematic diagram of the block antiferromagnetic
structure in the I4/m unit cell. 
Only Fe ions with their spin directions are shown. 
$J_c$ is the exchange interaction between 
spins in adjacent Fe planes. (b) Each shaded square highlights a block
of four ferromagnetically coupled Fe$^{2+}$ ions in the Fe plate. The open (orange) and the
crossed (blue) balls represent spins with opposite directions perpendicular to
the $ab$-plane. The black line marks the unit cell.  The four unique in-plane
exchange interactions considered in this work are labeled. 
(c) Theoretical spin wave dispersions calculated using experimentally determined
parameters.
}
\end{figure}

\begin{figure*}[bt!]
\label{fig2}
\includegraphics[scale=.94]{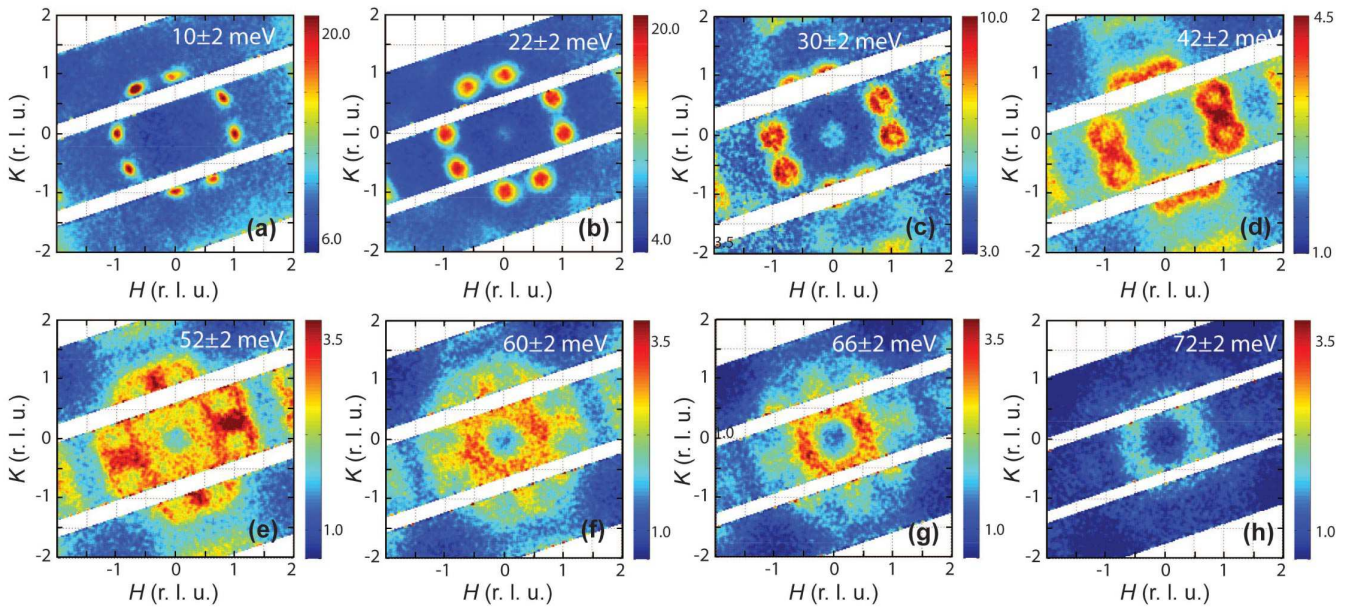}
\caption{(color online) Constant energy slices of the acoustic branch of the
spin wave excitations projected on the $(H,K,0)$ plane. The energy transfer is specified on each figure.
The relative intensity is indicated by the color scale. 
The sample was aligned on one of the two crystalline twins in the $I4/m$ unit cell.
The data were collected at SEQUOIA with $E_i = 50$ meV for (a-b) and 100 meV for
the other panels.  
 }
\end{figure*}

Here we report inelastic neutron scattering measurements covering the whole spin 
excitation spectrum in the (Tl,Rb)$_2$Fe$_4$Se$_5$ superconductor. 
Four doubly degenerate spin wave branches, one acoustic and three optical, 
form in three groups and span an energy range up to $\sim$210 meV [Fig.~1(c)]. 
A Heisenberg model, involving the intra- and inter-block
nearest and next nearest neighbor interactions in the Fe plane 
and the nearest neighbor interaction
between the planes, is sufficient to describe our data. 
Consistent with {\it ab initio} theoretical works, there is no need to resort to the
third nearest neighbor exchange interaction $J_3$.

Single crystals of (Tl,Rb)$_2$Fe$_4$Se$_5$  ($T_c \approx  32$ K) were grown 
using the Bridgman method \cite{D010462}.
A small single-crystal was used in a neutron diffraction study to 
determine the crystalline and magnetic structural properties \cite{D022882}.
For the inelastic studies, 240 plate-like crystals were mutually aligned on Al plates 
using an X-ray Laue diffractometer. The final assembly has a net sample mass of 19.5 g.
The sample was sealed with He exchange gas inside an Al can \cite{stone2011ultrathin} 
and measured with the SEQUOIA \cite{Granroth2006,Granroth2010}
fine resolution and the ARCS \cite{abernathy2012design} wide angular range Fermi chopper spectrometers 
at the Spallation Neutron Source (SNS) at Oak Ridge
National Laboratory (ORNL). 
Neutron beams of incident energy $E_i=50$, 100, 200 and 350 meV were
provided by the coarser resolution Fermi chopper \cite{Granroth2010} 
spinning at 180, 240, 360 and 420 Hz 
respectively on SEQUOIA. For ARCS $E_i=400$ meV was provided by the 700 meV Fermi chopper spinning 
at 420 Hz. 
The sample was kept in its ground state by 
a closed cycle refrigerator operating at $T\approx 6 K$. We will label the wavevector 
transfer ${\bf Q}=(H,K,L)$ using the tetragonal $I4/m$ unit cell \cite{D020830}
of $a=8.683$ and $c=14.39 \AA$.

The importance of the nearest and next nearest neighbor exchange interactions in the Fe plane 
was identified in the initial {\it ab initio} study examining the iron pnictide superconductors \cite{A042252}.
The lattice tetramerization forms the spin quartet block [Fig.~1(b)]
making the intra- and inter-block exchange interactions inequivalent \cite{D021344}.
Therefore, the effective Heisenberg Hamiltonian 
\begin{equation}
H =  \sum_{i,j} J_{i,j} S_{i} \cdot S_{j} - \Delta \sum_i S^2 _{iz}      
\label{}           
\end{equation}
is used which includes five exchange constants $J_1$, $J_2$, $J_1^{\prime}$, $J_2^{\prime}$ and $J_c$ 
as depicted in Fig.~1(a-b), and the single-ion anisotropy constant $\Delta$ that 
quantifies the observed Fe spin $S=3.2(1)/g$ alignment along the c-axis \cite{D022882}.
This spin model on the $\sqrt{5}\times\sqrt{5}$ vacancy ordered lattice has been theoretically investigated \cite{D033884,D041445},
and as to be shown later, describes the spin dynamics of (Tl,Rb)$_2$Fe$_4$Se$_5$.

Figure~2 shows the evolution of the acoustic branch of spin waves 
with increasing energy in (Tl,Rb)$_2$Fe$_4$Se$_5$.  
The orientation of the tetramer Fe block with respect to the $I4/m$ unit cell can be
clockwise [Fig.~1(b)] or counter-clockwise. The corresponding twins lead to eight,
instead of four, Bragg spots for the magnetic \{1,0,1\} peaks projecting on the $(H,K,0)$ plane. 
As energy transfer is increased, 
these spots broaden [Fig.~2(a-b)] and then develop into well-resolved circular rings above $\sim$ 30 meV [Fig.~2(c-g)]. The spin wave dispersions along the high symmetry direction [100] and
[110] are demonstrated in Fig.~3(a) and (b), respectively. There exists featureless scattering below 40 meV whose intensity increases with $Q$ and is attributed to multi-phonon scattering.

\begin{figure}[t!]
\label{fig3}
\includegraphics [width=\columnwidth]{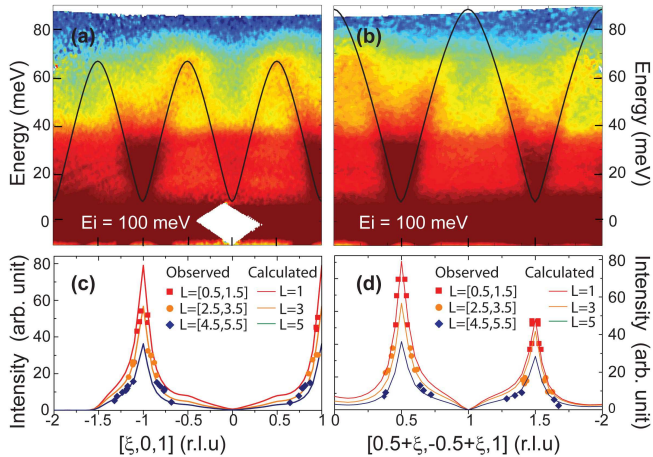}
\caption{(color online) Slices of the spin wave spectrum 
(a) along the [100] and (b) the [110] direction,
measured with $E_i= 100$ meV. 
The solid line is the theoretical spin wave dispersion described in the text. 
The spectral weight of the acoustic branch
along the (c) [100] and (d) [110] direction was obtained from the
constant-$E$ cuts at different $L$ values.
The solid lines are calculated intensities at corresponding $L$. 
}
\end{figure}

Across an energy gap, and above the acoustic branch, are two optic branches. Another
gap proceeds a third optic branch at higher energy.
The projection of the optical modes at 110$\pm$3 meV and 205$\pm$15 meV on the $(H,K,0)$ plane
are shown in Fig.~4(a) and (c), respectively. 
To obtain eigenvalues of the optic modes at high
symmetry points, constant-${\bf Q}$ cuts at peak and background positions were
performed and their difference was fit to a Lorentzian. Example curves and fits are shown in
Figs.~4(b) and (d). The peak at $E= 209(1)$ meV in Fig.~4(d) presents 
the highest energy magnetic excitation mode in (Tl,Rb)$_2$Fe$_4$Se$_5$.

\begin{figure}[t!]
\label{fig4}
\includegraphics[width=0.9\columnwidth]{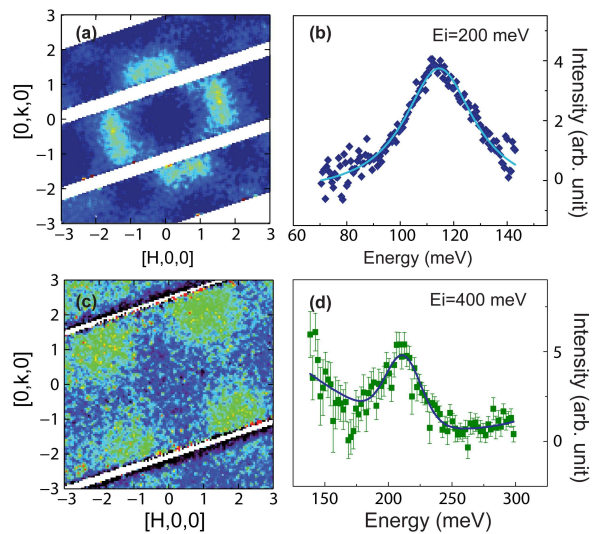}
\caption{(color online) Constant energy slices of the optic branches of the
spin wave excitations projected on the $(H,K,0)$ plane at the energy transfer of (a)
$E=107-113$~meV and (c) $E=190-220$~meV. 
(b) The constant-${\bf Q}$ cut at (1.5,0.5,0) with a background at (3,-1,0) subtracted. 
(d) The constant-${\bf Q}$ cut at (2,1,0) with a background at (3,1,0) subtracted.}
\end{figure}

At the low energy end, the single-ion anisotropy $\Delta$ in Eq.~(1) 
breaks the Heisenberg spin rotation symmetry, thus, opening a gap in 
acoustic spin waves at magnetic Bragg points.
The inter-plane coupling $J_c$, which stabilizes the antiferromagnetic order
at finite temperature, also introduces a modulation in spin wave dispersion
along the $c$-axis. 
Fig.~5(a) and (b) show the details of the low-energy spin
wave excitations obtained with the finer $E$-resolution spectrometer configuration of $E_i=50$ meV.
The energy gap in magnetic excitations is obvious.
Constant-${\bf Q}$ cut through the magnetic zone center (background) is shown in Fig.~5(c). 
The difference intensity was fit to a 
step function convoluted with the instrument resolution to obtain the intrinsic gap value 6.5(3) meV. 
In Fig.~5(b), the dispersive curve of bandwidth $\sim$18 meV along the $c$-axis is
clearly observed.

The simultaneous fit of the data from all branches along the multiple symmetry directions
measured in this experiment to the spin wave solution of Eq.~(1) yields the parameters: 
\begin{align}
SJ_1	&=	-30(1)  {\rm meV}, & SJ_1^{\prime} &= 31(13)  {\rm meV}, \nonumber \\ 
SJ_2&=10(2)   {\rm meV}, & SJ_2^{\prime} &= 29(6)  {\rm meV},\nonumber \\ 
SJ_c&=0.8(1)  {\rm meV}, & S\Delta &= 0.3(1)   {\rm meV}.
\end{align}
  The resulting spin wave dispersion curves in various high symmetry directions are shown in
Fig.~1(c). They are also reproduced as the solid
lines in Fig.~3(a-b) and Fig.~5(a-b), and are in excellent agreement with the
measurements. To further check the reliability of the fits,
the inelastic neutron scattering intensity 
was also calculated using these fitting parameters and over-plotted
with the observed intensity in Figs.~3(c) and (d). The theory agrees well with 
experimental results.  

\begin{figure}[t!]
\label{fig5}
\includegraphics[width=\columnwidth]{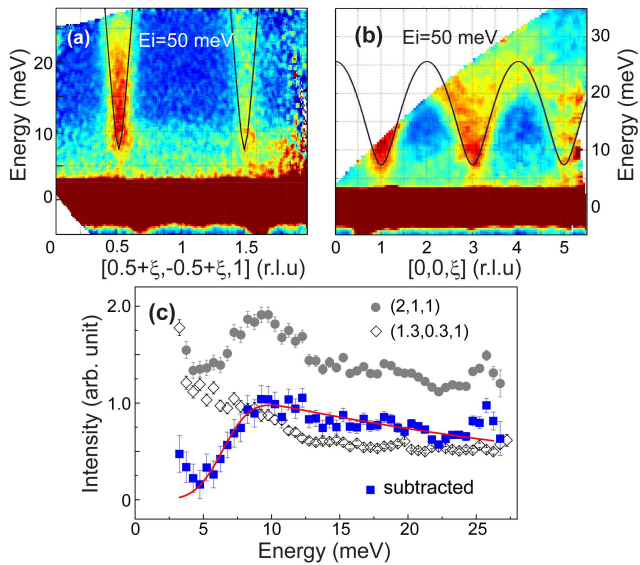}
\caption{(color online) Slices of the acoustic spin wave branch along  
(a) the [110] and (b) the [001] direction.  (c) Constant-${\bf Q}$ cuts through
the magnetic Bragg point (2,1,1) and the background point (1.3,0.3,1). The difference
intensity curve was fitted to the spin wave excitations convoluted with the instrumental resolution
to obtain the energy gap at 6.5(3) meV.
}
\end{figure}

The $J_1$, $J'_1$, $J'_2$, $J_c$ and $\Delta$ in Eqs.~(2) have the correct sign to stabilize 
the observed block antiferromagnetic order, while the weaker antiferromagnetic 
$J_2$ frustrates the ferromagnetically aligned spin block [Fig.~1(b)].
The strong difference of the exchange constants between the intra and inter-block nearest
neighbor and next nearest neighbor Fe spin pairs highlights the electronic consequence
of the lattice tetramerization in the $\sqrt{5}\times\sqrt{5}$ structure 
uncovered in structural refinement 
studies \cite{D020830,D014882} and emphasized by electronic structure calculations 
\cite{D021344,E056404,E060881}. In particular, the recent {\it ab initio} linear response theory
concludes that the significant in-plane exchange interactions include only 
$J_1$, $J_2$, $J'_1$ and $J'_2$,  whose calculated values \cite{E056404}, remarkably, 
agree with our experimental results in Eqs.~(2) qualitatively. 

The block antiferromagnetic order not only exists in superconducting samples, but also in
insulating samples with a less ordered $\sqrt{5}\times\sqrt{5}$ vacancy structure.
Experimentally this type of order is observed even for 29\% filling of the vacant Fe site \cite{D023674} and {\it ab initio}
calculations at 25\% filling support the stability of the block antiferromagnetic order \cite{E056404}.
Recently, the spin dynamics in an insulating Rb$_{0.89}$Fe$_{1.58}$Se$_2$ sample with
the block antiferromagnetic order were investigated with inelastic neutron 
scattering \cite{E054675}. The overall energy scale of spin dynamics is similar to the 
system under study in this paper. 
However the published analysis of that data provides a somewhat anomalous result as compared to our result and the aforementioned {\em ab initio} works. Specifically their data could be least-squared fit only by a model that includes an inter-block $J_3$ term in Eq.~(1), where the preponderance of other results finds no need for this term.  This discrepancy can be reconciled in two ways: One is to introduce a subtle physical effect in the {\em ab initio} studies that makes the insulating sample subtly different from the superconducting sample. The other is to consider a small overlap of the observed excitation with the {\bf Q}-resolution tail of its twin. We minimized this latter effect in our data by using the fine {\bf Q} resolution of SEQUOIA for our acoustic mode measurements.

In addition to the observed block antiferromagnetic order \cite{D020830,D022882,D023674}, 
many other magnetic order configurations are possible with the Heisenberg model
Eq.~(1) with in-plane exchange interaction extending only to the next nearest neighbors
on the $\sqrt{5}\times\sqrt{5}$ vacancy ordered lattice. Theoretical work has investigated 
the stability of these spin configurations \cite{D021344,D033884,D041445}. According to the calculated
phase diagram \cite{D041445}, the exchange constants determined in the present study
put (Tl,Rb)$_2$Fe$_4$Se$_5$ near the boundary between the block antiferromagnetic phase
and a non-collinear antiferromagnetic phase. Furthermore, such a
non-collinear antiferromagnetic order has recently been observed in a spin-flop
transition from the block antiferromagnetic order at 100 K in TlFe$_{1.6}$Se$_2$
(or Tl$_{2.5}$Fe$_4$Se$_5$) \cite{E071318}. Therefore we postulate that composition tuning
has pushed Tl$_{2}$Fe$_4$Se$_5$ just across this phase boundary.

The success of the Heisenberg localized spin model in describing spin dynamics in
the (Tl,Rb)$_2$Fe$_4$Se$_5$ superconductor may be attributed to the fact that
the observed saturated magnetic moment 3.2(1)$\mu_B$/Fe is very large \cite{D022882}. 
Therefore the system is close to the local spin limit. In the iron chalcogenide Fe$_{1+y}$Te$_{1-x}$Se$_x$ superconductors
without the intercalating layer and Fe vacancies, 
the spin excitations belong to the itinerant 
class as those in the antiferromagnet Cr 
which are described by more complex theories than linearized spin-wave theory \cite{B092417,B114713}.
Even-though, the binary iron chalcogenides are not simple antiferromagnetic metal like Cr: Their parent compounds 
have a magnetic wave vector that cannot be accounted for by Fermi surface nesting, and have the largest saturated magnetic 
moment (2$\mu_B$/Fe) among the previously discovered families of Fe-based superconductors \cite{A092058}. 

Moreover,
a diffusive spin excitation component in the binary iron chalcogenides was recently resolved as originating from 
interstitial Fe-induced short-range spin plaquettes that contain the same four-spin blocks
as found in the block antiferromagnetic order \cite{D095196}. Such fluctuating spin quartets 
have been shown to contribute pronounced
features in the spin dynamics in the parent compound FeTe$_{1.1}$ \cite{D035073}.
The close link among these antiferromagnetic states discovered experimentally in iron chalcogenides
with or without vacancies, was
anticipated in a theory including both itinerant and localized electronic states \cite{E060881}.

In summary, in this inelastic neutron scattering work, we contribute fresh insights to 
the understanding of iron chalcogenide superconductors by determining the spin Hamiltonian 
for a new block antiferromagnetic
order in the (Tl,Rb)$_2$Fe$_4$Se$_5$ superconductor. Our results agree with the majority
of theoretical studies that state that the dominant exchange interactions extend only to the next nearest neighbor
Fe pairs in the plane. The block antiferromagnetic order is frustrated only by the intra-block
next nearest exchange $J_2$, the weakest among the four dominant in-plane exchange interactions.
Combining our experimental exchange parameters with theoretical calculations shows that this system  is near the boundary of the block antiferromagnetism regime.
A unified theoretical framework for all observed types of magnetic order in the Fe-based
superconductors should progress based on the experimental findings presented here.

We thank A.~I.\ Kolesnikov, C.\ Cao, J.\ Dai and D.-H. Lee for fruitful discussions.  
The works at RUC and ZU were supported by the National Basic Research Program of China Grant Nos.\ 2012CB921700, 2011CBA00112, 
2011CBA00103, 2012CB821404 and 2009CB929104 and by 
the National Science Foundation of China Grant Nos.\ 11034012, 11190024, 10974175 and 11204059.
This Research at Oak Ridge National Laboratory's Spallation Neutron Source was sponsored by the Scientific User Facilities Division, Office of Basic Energy Sciences, U. S. Department of Energy.


\end{document}